\documentclass{article}

\begin{document}

\def\giorno{1 November 2001}

\def\pa{\partial}
\def\a{\alpha}
\def\b{\beta}
\def\ga{\gamma}
\def\eps{\eta}
\def\la{\lambda}
\def\s{\sigma}
\def\om{\omega}

\def\E{{\cal E}} 
\def\F{{\cal F}}
\def\L{{\cal L}}
\def\M{{\cal M}}
\def\Y{{\cal Y}}
\def\X{{\cal X}}
\def\V{{\cal V}} 
\def\W{{\cal W}}
\def\Y{{\cal Y}}
\def\H{{\cal H}}
\def\h{{\cal H}}
\def\G{{\cal G}}

\def\T{{\rm T}}

\def\sse{\subseteq} 
\def\ss{\subset}

\def\ker{{\rm Ker}} 
\def\ran{{\rm Ran}} 

\def\({\left(}
\def\){\right)}
\def\[{\left[} 
\def\]{\right]} 

\def\~#1{{\widetilde #1}}
\def\^#1{{\widehat #1}}
\def\=#1{{\widetilde #1}}
\def\frac#1#2{{#1 \over #2}}

\def\hot{{\rm h.o.t.}}

\def\eb{{\bf e}}
\def\pb{{\bf p}}
\def\vb{{\bf v}}
\def\xb{{\bf x}}

\def\C{{\bf C}}
\def\N{{\bf N}}
\def\R{{\bf R}}
\def\Q{{\bf Q}}
\def\cR{{\cal R}}

\title{{\bf Resonant normal forms \\ as constrained linear systems}}

\author{Giuseppe Gaeta\footnote{Supported by ``Fondazione CARIPLO per la ricerca scientifica'' under project ``Teoria delle perturbazioni per sistemi con simmetria''; E-mail: gaeta@berlioz.mat.unimi.it or gaeta@roma1.infn.it}  \\ 
{\it Dipartimento di Matematica, Universit\'a di Milano,} \\
{\it V. Saldini 50, I--20133 Milano (Italy)} }

\date{\giorno}

\maketitle
 
\noindent {\bf Summary.} We show that a nonlinear dynamical system in Poincar\'e-Dulac normal form (in $\R^n$) can be seen as a constrained linear system; the constraints are given by the resonance conditions satisfied by the spectrum of (the linear part of) the system and identify a naturally invariant manifold for the flow of the ``parent'' linear system. The parent system is finite dimensional if the spectrum satisfies only a finite number of resonance conditions, as implied e.g. by the Poincar\'e condition. In this case our result can be used to integrate resonant normal forms, and sheds light on the geometry behind the classical integration method of Horn, Lyapounov and Dulac.

\bigskip\bigskip

\section*{Introduction}

Normal forms are central to our understanding of nonlinear dynamics around known solutions, in more ways than we could recall here; see e.g. \cite{Arn1,Arn2}. 

In this note we will consider dynamical systems -- or, equivalently, vector fields -- in $\R^n$ which are in normal form around a regular critical point (for normal forms in this context, see \cite{Arn1,Arn2,Elp,IAd,Wal}, or the short introductions given e.g. in \cite{CGs,Gle,GuH,Ver}).

We recall that if the linear part of the system is nonresonant (this and other relevant definitions will be recalled in section 1), the normal form is linear; we will thus assume the linear part is resonant. If the system is in normal form, all nonlinear terms will be resonant.

We will show that the time evolution of resonant nonlinear terms can always be written as a linear automorphism of their set; this will allow to map the nonlinear system to a ``parent'' system of larger dimension, but {\it linear}. 

If the linear part satisfies some additional conditions -- in particular, the Poincar\'e condition, i.e. if the spectrum of the linear part lies in a Poincar\'e domain -- the linear system has finite dimension, and the construction considered here leads to an explicit and simple method of integration. 

Resonant normal forms satisfying the Poincar\'e condition are well known to be integrable using a ``triangular'' property of equations in resonant normal form which was already remarked by Dulac \cite{Dul}, and which guarantees integrability also with a classical procedure (Dulac attributes this to Horn and Lyapunov); for related modern ideas see \cite{Wal}. The present note shows how this very classical result is related to the possibility of writing a resonant normal form as a constrained linear system.

Actually, this note implements for resonant normal forms ideas which were put forward (in the hamiltonian context) by several authors, in particular Kazhdan, Kostant and Sternberg \cite{KKS} (see also \cite{Mle,Mar}).

Their point is, roughly speaking, that while in many cases one uses a symmetry quotient (Marsden-Weinstein reduction) to reduce a nonlinear system to a lower dimensional one, and maybe integrate it, in other cases the nonlinear system can be seen as the reduction to a lower dimensional invariant manifold of a linear system living in a higher dimensional space, and can be integrated by ``enlargement'' rather than by reduction. 

This mechanism is at work in systems of Lax type (e.g. Calogero-type systems), where an $n$-dimensional nonlinear system is embedded into a higher dimensional -- typically $n^2$ dimensional, resorting to $(n \times n)$ matrices -- linear system. It was even conjectured \cite{Mar} that all integrable systems actually originate in higher dimensional linear systems, which we observe only through their projection to a lower dimensional nonlinear manifold. 

The present work shows that the Kazdan-Kostant-Sternberg mechanism can work also for non-hamiltonian systems, and that it applies in particular to resonant normal forms.

\medskip

After a preliminary version of this note was circulated, prof. Walcher pointed out that the basic ideas behind the treatment of resonant  normal forms given here can also be found in \cite{Wal}; however I believe that the implementation given in this note is somewhat simpler, and clarifies the geometry involved in the problem and in the integration method.

\subsection*{Acknowledgements} 

This work came out of an attempt to apply to resonant normal forms the  geometric ideas on integrable systems originating in \cite{KKS}, first exposed to the author by Giuseppe Marmo a long time ago; it was also triggered by questions posed by Dario Bambusi. I would like to warmly thank both of them, for this and many other things. 
The financial support of ``{\it Fondazione CARIPLO per la ricerca scientifica}'' under the project {\it ``Teoria delle perturbazioni per sistemi con simmetria''} is gratefully acknowledged.

\section{Normal forms}

In this section we fix notation and collect several definitions, and properties of normal forms, to be used in the following.
\medskip

Let us consider a $C^\infty$ function $f: \R^n \to \R^n$ such that $f(0) =0$, expanded in a series of homogeneous terms $f_k (x)$, where  $f_k (a x) = a^{k+1} f_k (x)$ for all real number $a$. 
This defines a dynamical system ${\dot x} = f(x)$ in $V_0 = \R^n$ having a critical point at the origin; we also rewrite this, using the expansion in homogeneous terms and singling out the linear part, as
$$ {\dot x} \ = \ Ax \ + \ F (x) \ = \ Ax \ + \ \sum_{k=1}^\infty f_k (x) \ . \eqno(1)$$
We will assume that the matrix $A:= (Df)(0)$ is in Jordan form.  

\medskip
{\bf Remark 1.} Transforming a general matrix to Jordan form can be, in practice, very hard for large $n$; see \cite{ScW} for a discussion of this point in the context of normal forms theory. $\odot$
\medskip

We consider several vector fields associated to the system and to the decomposition given in (1), to be used in the following; we always write $\pa_i$ for $\pa / \pa x^i$. As well known the matrix $A$ can be decomposed into a semisimple and a nilpotent part, mutually commuting, which we denote as $A_s$ and $A_n$; as $[A_s , A_n ] =0$, we also have $[A_s , A] = 0 = [A_n , A]$.  

We will use $X_f = f^i (x) \pa_i$, $X_A = (Ax)^i \pa_i$, $X_0 = (A_s x)^i \pa_i$, and $X_F = F^i (x) \pa_i$; we also consider $X_\ell = (A^+ x)^i \pa_i$; as long as we deal with real matrices $A$, this reduces to $X_\ell = (A^T x)^i \pa_i$. Note that $[ X_0 , X_A ] = 0 = [X_0 , X_\ell ]$.

Finally, we recall that a matrix $A$ is said to be normal if it commutes with its adjoint, $[A,A^+] =0$; obviously this is equivalent to the condition that $[ X_A , X_\ell ] = 0 $. In general this is not satisfied, and Jacobi identity only guarantees that $[ X_0 , [X_A , X_\ell ] ] = 0$.

\medskip

Let $(\la_1 , ... , \la_n )$ be the eigenvalues of $A$ (we denote by $\s$ their ensemble, i.e. the spectrum of $A$), and let $(\eb_1 , ... , \eb_n )$ be the basis in $V_0 = \R^n$ with respect to which the $x^i$ coordinates are defined. 
We will use the multiindex notation $$ x^\mu \ := \ x_1^{\mu_1} ... x_n^{\mu_n} \ .$$ 
Then the vector $\vb_{(\mu)} := x^\mu \eb_\a$ is resonant with $A$ (or {\it resonant } for short) if 
$$ (\mu \cdot \la ) \ := \ \sum_{i=1}^n \ \mu_i \, \la_i \ = \ \la_\a \ \ \ {\rm with} \ \ \mu_i \ge 0 \ , \ \ |\mu| := \sum_{i=1}^n \mu_i \ge 2 \ . \eqno(2)$$
Note that when we define and determine resonant monomials and vectors, we can as well consider $A_s$ rather than $A$. 

The space of vectors resonant with (the semisimple part of) $A$ is defined as the linear span of such vectors; we will consider a basis $(\vb_1 , ... , \vb_r )$ in this. Thus $F$ is resonant if and only if $F = c_i \vb_i$ for some constants $c_i$. 

\medskip
{\bf Remark 2.} We stress that $r$ could be infinite, but will always be finite if $A$ admits a finite number of resonance relations, e.g. if $\s$ satisfies the Poincar\'e condition. On the other side, if the $\la_i$ satisfy a ``master resonance'' relation, i.e. $\sum_i \mu_i \la_i = 0$ with $\mu_i$ as in (2) and $|\mu| > 0$ (notice in this case $\s$ cannot satisfy the Poincar\'e condition), then there is an infinite number of resonances. This is in particular the case if there is a zero eigenvalue or if we deal with a resonant Hamiltonian system. $\odot$
\medskip

A monomial $x^\mu$ such that (2) is satisfied for some $r \in (1,...,n)$ is called a {\it resonant monomial}. We consider a basis of resonant monomials $\{ \phi^1 (x) , ... , \phi^r (x) \}$, and their linear span; this is a linear vector space $V_1$ (in the space of scalar polynomials on $\R^n$). We choose a basis $\{ \pb_1 , ... , \pb_r \}$ in this. Here $\pb^i$ corresponds to $\phi^i (x)$, i.e. the scalar polynomial $\sum_{i=1}^r c_i \phi^i (x) $ is represented in $V_1$ by the vector $\sum_{i=1}^r c_i \pb_i$.

\medskip

We say (see e.g. \cite{Elp}) that (1) is in Poincar\'e-Dulac normal form if the vector fields $X_\ell$ and $X_F$  commute: 
$$ \[ \, X_\ell \, , \, X_F \, \] \ = \ 0 \ . \eqno(3)$$
This implies that all nonlinear terms are resonant with $A$ (i.e. with $A_s$); however not all resonant terms will satisfy (3) when $A_n \not= 0$, see e.g. example 3 below.

Notice that in general $[X_\ell , X_A ] \not= 0$: thus, unless $A$ is normal, we cannot affirm that $X_\ell$ (or $X_A$) commutes with $X_f$. 

However it is easy to see from (3) that for systems (1) in Poincar\'e-Dulac normal form, both $X_A$ and $X_F$ commute with $X_0$, and therefore
$$ \[ \, X_0 \, , \, X_f \, \] \ = \ 0 \eqno(4)$$

When the system satisfies (4) -- although (3) is possibly not satisfied -- i.e. if $F$ is resonant with $A_s$, we say that it is in normal form with respect to $A_s$ or, that it is in {\it seminormal form}.

As well known, starting from any dynamical system (or vector field)  of the form (1), we can arrive to a dynamical system (or vector field) in Poincar\'e-Dulac normal form by means of a sequence (in general, infinite) of near-identity transformation obtained by means of the Poincar\'e algorithm; these combine into a near-identity transformation $H$ defined by a series which is in general only formal. The same applies for seminormal forms.

However, one can guarantee the convergence of the series on the basis of properties of the spectrum $\s$ of the linear part $A$. In particular, it was already known to Poincar\'e and Dulac that convergence is guaranteed if the convex hull of $\s$ in the complex plane does not include the origin; in this case we say that $\s$ belongs to a Poincar\'e domain, or that $A$ satisfies the Poincar\'e condition, or also that $\s$ is a Poincar\'e spectrum. This relevant property makes the study of normal forms with Poincar\'e spectrum specially interesting.

It is easy to see -- and again was well known to Poincar\'e and Dulac -- that if $\s$ satisfies the Poincar\'e condition, then only a finite number of resonances is present, i.e. the Poincar\'e-Dulac normal form is finite (conditions ensuring finiteness of the normal form are discussed in \cite{GaW}). 

\medskip
{\bf Remark 3.} Other conditions guaranteeing convergence of $H$, on the basis of $\s$ and of symmetry properties of the normal form, are also known, see e.g. the review \cite{CWspt}; in such cases one is not guaranteed to have an integrable normal form and is thus no surprise that the method presented here does in general not apply. $\odot$

\section{Normal forms as reduction of a linear system}

In this section we will give a very elementary procedure to associate to a nonlinear system (1) in resonant normal form a linear system ${\dot \xi} = B \xi$ in a real vector space $V \simeq V_0 \oplus V_1$ (this can be seen as a trivial bundle $\pi: V \to V_0$ over $V_0$) and a set of constraints ${\cal E}$ so that $\E = 0$ identifies a smooth (actually, algebraic) submanifold $\M \ss V$, invariant under the flow of the linear system. The resonant normal form is just the reduction of the linear system to $\M$. 

If the normal form is finite then $V = \R^N$ for some finite $N > n$, and this procedure also provides a way to explicitely and elementarily integrate the system in normal form, as discussed in the section 3.

\subsection{General construction}

Let $V_1$, $\phi_i (x)$ and $\pb_i$ be as defined above; we assume for ease of language that $r$ is finite (as mentioned above, this is the case for $\s$ satisfying the Poincar\'e condition). We consider the real vector space $V = V_0 \oplus V_1 = \R^{n+r}$ with basis vectors  $(\eb_1 , ... , \eb_n ; \pb_1 , ... , \pb_r )$ and coordinates $\xi = (x^1 , ... , x^n ; w^1, ... , w^r)$. 

We consider a ``parent'' dynamical system ${\dot \xi} = \psi (\xi)$ [a vector field $Y_\psi = \psi^j (\xi) (\pa / \pa \xi^j)$~] in $V$  defined as follows: first we rewrite (1) substituting $w^1 , ... , w^r$ for $\phi^1 (x) , ... , \phi^r (x)$ (this gives the evolution equation for the $x$'s); then we assign time evolution for the $w$ by $d w^i / dt = (\pa \phi^i (x) / \pa x^j) (d x^j / dt )$. Having written these equations, we will now consider the $x$ and $w$ as indipendent quantities. 

It is clear that, by construction, the manifold $\M \ss V$ defined by 
$$ \E^i \ := \ w^i - \phi^i (x^1 , ... , x^n) = 0 \ \ \ \ \forall i = 1,...,r \eqno(5)$$
is invariant under the flow of the system we have defined in this way, i.e. $Y_\psi : \M \to \T \M$. 

The $\M$ defined by (5) is an algebraic manifold (since the $\phi$ are polynomials) and is tangent in the origin to the linear space $\R^n$ defined by $w^i = 0$, i.e. to $V_0 \ss V$ (since the $\phi$ are nonlinear functions of the $x$). It is also obvious from (5) that it is a global section of $\pi: V \to V_0$. Thus a smooth dynamics on $\M$ projects globally to a smooth dynamics in $V_0$. 

On the invariant manifold $\M$, the parent system is equivalent by construction to the original one. 

\medskip
{\bf Remark 4.} In the fiber bundle notation, and denoting by $\pi^*$ the lift of $\pi: V \to V_0$ to $\pi^* : \T V \to \T V_0$ and by $Y_\psi^\M$ the restriction of $Y_\psi$ to $\M$, we have $\pi^* [Y_\psi^\M] = X_f$.
$\odot$ \medskip

We will show in the next subsection that {\it the evolution equation we obtain for $\xi:=(x,w)$ is linear}, i.e. ${\dot \xi} = \Psi (\xi)$ reduces to ${\dot \xi} = B \xi$, with $B$ a matrix.

The evolution equations constructed in this way have also several other general features in common; we mention these omitting the elementary proof.

(1) We actually obtain a ``block triangular'' evolution equation: the evolution of the $w$ depends only on the $w$ themselves, while that of the $x$ depends on the $x$ and the $w$ together.

(2) The system of ODEs we obtain is also, if the coordinates are properly arranged\footnote{in general, we can have to pass to complex coordinates in the $x$ space and to have a change of variables in the $w$ space to obtain this, see example 4}, triangular in proper sense. 

(3) The eigenvalues of $B$ (see point 1) are given by $ (\la_1 , ... , \la_n ; \la_{\a_1} , ... , \la_{\a_r} )$, where the $\la_i$'s are the eigenvalues of $A$, and $\a_i$ is the $\a$ associated to the resonant monomials $\phi_i$, see (2); thus, we always have multiple eigenvalues.

\subsection{Proof of the linearity of the parent system}

We will now substantiate the assertion that the parent system obtained according to the procedure given above will always be linear. We will actually proof this in two ways, i.e. both algebraically and geometrically.
\bigskip

\noindent
{\bf Algebraic proof.}

Summation over repeated indices will be understood, and we will denote by $\nu(i)$ the multiindex such that $\nu_j = \delta_{ij}$.

Let us consider a resonant monomial $w = x^\mu$, and say it satisfies $(\la \cdot \mu) := \la_i \mu_i = \la_\a$. We denote by $w_{\a;i;\s}$,  $i=1,...,q(\a)$ the resonant monomials $x^\s $ such that $(\la \cdot \s) = \la_\a$ (with the same fixed $\a$).
Then we have that (for $f$ in seminormal form) under ${\dot x} = f(x)$ the time evolution of $w$ is given by $X_f (w)$, i.e. 
$$ {d w \over d t} \ = \ {\pa w \over \pa x^i} \ f^i (x) \ = \ \mu_i x^{\mu - \nu(i)} \ [ (A_s)^i_{\, j} x^j + (A_n)^i_{\, j} x^j + c_m w_{\a;m;\s}  ] \ , $$
where we have used the decomposition (1) and the fact all the nonlinear terms must be resonant. 

We assume that $A$ is in Jordan form, so that $A_s = {\rm diag} (\la_1 , ... , \la_n)$, and $(A_n)^i_{\, j} = \eta_{ij}$ is different from zero (and equal to one) if and only if $j=i+1$ and $x^i,x^j$ belong to the same Jordan block; this implies of course that $\la_i = \la_j$.

Notice that terms with $\mu_i = 0$ are absent from the sum over $i$; we can therefore assume $\mu_i \not= 0$. Under this condition, and using the assumption that $A$ is in Jordan form, we can rewrite  
$$ {\dot w} \ = \ (\la \cdot \mu) \, x^\mu \ + \ \mu_i \, \eta_{ij} \, x^{\mu - \nu(i) + \nu(j)} \ + \ c_\s \, \mu_i \, x^{\mu - \nu(i) + \s} \ . \eqno(6)$$
The first term on the r.h.s. is nothing else that $\la_\a w$. We want to check that the other terms are also (the sum of) resonant monomials; in order to do this we do not have to worry about the scalar coefficients in fronts of them.

The monomials appearing in the second term of the r.h.s. are of the form $x^\varphi = x^{\mu - \nu(j) + \nu(k)}$, and we can assume $\mu_i \not= 0 $ and $\eta_{ij} \not= 0$ (or the corresponding monomial would not be present in the sum). For these we have
$$ (\la \cdot \varphi) \ := \ \la_i \varphi_i \ = \ (\la \cdot \mu ) - \la_i + \la_j \ = \ (\la \cdot \mu ) \ = \ \la_\a \ ; $$
we have used the fact that $\eta_{ij} \not= 0$ implies $\la_i = \la_j$, and the resonance relation satisfied by $w$ itself. Thus the second term in the r.h.s. of (5) is the sum of resonant monomials (with the same $\a$ as $w$). 

The monomials appearing in the third term are of the form $x^\varphi = x^{\mu - \nu (i) + \s}$, where $(\la \cdot \s) = \la_i$ and we can assume $\mu_i \not= 0$ (or the corresponding monomial would be absent from the sum). We have now
$$ (\la \cdot \varphi) \ := \ \la_i \varphi_i \ = \ (\la \cdot \mu ) - \la_i + (\la \cdot \s) \ = \ \la_\a \ ; $$
again we have a sum of resonant monomials (with the same $\a$ as $w$).

This concludes the proof that the right hand side of (6) can be written as a linear combination of resonant monomials, i.e. that the evolution equation constructed according to our procedure is linear. \hfill $\triangle$
\bigskip

{\bf Remark 5.} Notice that we have actually proved something more, i.e. that if $w = x^\mu$ with $(\la \cdot \mu) = \la_\a$, only resonant monomials $\=w = x^\pi$ with $(\la \cdot \pi) = \la_\a$ (with the same $\a$ as above) will appear in this linear combination. That is, the matrix $B$ will be a block one, where the blocks correspond to resonant monomials identified as described here. $\odot$
\bigskip

\noindent
{\bf Geometric proof}

A more geometric (but equivalent) proof could be obtained by considering a basis of (nonlinear) resonant vectors $x^\mu \eb_k$ ($k=1,...,n$) and the corresponding vector fields $X^{(i)} = (x^\mu) \pa_k$ ($i=1,...,r$, see above). These obviously generate a Lie algebra $\G$ (recall that the resonance condition is equivalent to commutation with $X_0$, and notice that the commutator of two vectors built from nonlinear monomial terms will never be a linear vector field). 

One should then check that $\G$ is globally invariant under time evolution, i.e. that $[X_f , \G ] \sse \G$.

Take $X_\phi \in \G$: we have $[X_f , X_\phi ] = [X_A , X_\phi ] + [X_F , X_\phi]$, and the second term is by definition in $\G$. To see that the first is also in $\G$, we have to check the vanishing of  $[X_0 , [X_A , X_\phi]]$; using the Jacobi identity, this is $[X_A,[X_0,X_\phi]] - [X_\phi , [X_0 , X_A]]$, and both terms vanish separately. The proof is complete. \hfill $\triangle$
\bigskip

{\bf Remark 6.} Notice that if $A$ is not normal, the set of resonant monomials defining vectors in normal form with respect to $A^+$ would not, in general, be closed under time evolution. This is also immediately seen from the alternative geometrical proof by remarking that if we substitute $X_\ell$ for $X_0$, we have (for $A$ not normal) $[X_\ell , X_A ] \not= 0$. Thus for non-normal $A$ we cannot limit to consider vectors in normal form, but have to consider the set of all resonant vectors. This will also be clearly shown in example 3 below. $\odot$

\subsection{Truncated normal forms}

When the normalizing transformation is not convergent, the normal form is not conjugated to the original system. However, in such a case one can consider normalization only up to a sufficiently low degree $N$ (in practice this is determined by either the computational limits or the optimal degree on the basis of convergence pèroperties of the truncated series); in this way one obtains a system which is in normal form up to order $N$. This system can be truncated at order $N$ -- thus obtaining a truncated normal form -- and the relation of such a truncation with the full system will then be studied via other techniques in perturbation theory \cite{Arn1,Arn2,Gle,Ver}. 

The block structure of the $B$ matrix, as determined above (see in particular remark 5), explains when this truncation will result in a closed parent system. 

Indeed, consider all the resonance relations (2); let $m_- (\a)$ and $m_+ (\a)$ be the smaller and greater values of $|\mu|$ for which a relation $(\mu \cdot \lambda ) = \la_\a$ is satisfied. 
It follows from remark 5 that if 
$$ N \ \not\in \ [m_- (\a) , m_+ (\a) ] \ \ \ \  \forall \, \a \ = \ 1 ,..., n \ \ , \eqno(7) $$ 
then the truncated normal form at order $N$ is mapped, by the procedure discussed in this note, to a closed linear system 
(note that for a finite dimensional system we always have a finite number of resonances with $|\mu| \le N$, for any finite $N$). 

Thus the integration procedure discussed here is also of use in cases where the normalizing transformation is not convergent in any neighbourhood of the origin. 

On the other hand, it should be stressed that if the system admits an infinite number of resonances, then (7) can be satisfied only for $N < m_- (\a )$ for all $\a$, but in such case the truncated normal form is trivial (it reduces to the linear part of the system). Thus, the truncated normal form will not result in a closed finite dimensional linear parent system, see example 4 below. Such a situation is met  when a ``master resonance'' is present (see remark 2), and in particular when dealing with resonant hamiltonian systems. 

\section{Integration of normal forms}

The strategy to integrate normal forms via the parent linear system ${\dot \xi} = B \xi$ we have defined is rather obvious: this rests on the dynamical invariance of the manifold defined by (5) and can be divided into three steps. That is, 

\begin{enumerate}

\item {\it Step 1.} For $\xi = (x;w)$, determine the general solution of the linear equation ${\dot \xi} = B \xi$ in $V = \R^{n+r}$, say with solution $ \^\xi (t)$ where $\^\xi (0) = \xi_0 = (x_0 , w_0)$ is the initial datum. This will depend on the $n+r$ arbitrary constants $(x_0,w_0)$.

\item {\it Step 2.} Restrict the general solution to the invariant $n$-dimensional submanifold $\M \ss \R^{n+r}$ defined by $w^i = \phi_i (x^1, ... , x^n)$. This will depend on the $n$ arbitrary constants $x_0$.

\item {\it Step 3.} Project the general solution $( x(t),w(t) )$ on $\M \ss V$ to the subspace $V_0 = \R^n$ spanned by the $x$ variables, i.e. extract $x(t)$ forgetting about $w(t)$.

\end{enumerate}

As discussed above, the correspondence between the original nonlinear system and the restriction of the parent system to the invariant manifold $\M$ is guaranteed by construction, and projection is globally well defined as $\M$ is identified by the algebraic equations (5). It is therefore clear that this procedure will indeed provide the most general solution to the original nonlinear system in $V_0$.

This strategy will be particularly simple, and successful, when there is only a finite number of resonances, and in particular when $\s$ belongs to a Poincar\'e domain.

\bigskip

It should be stressed that if $\s$ belongs to a Poincar\'e domain, the normal forms could of course also be integrated directly: indeed the corresponding system is nonlinear but, as remarked by Dulac \cite{Dul}, always in triangular form (cf. the properties of the parent system mentioned at the end of subsection 2.1). Namely, we can always write 
$ {\dot x}^i = A^i_{\, j} x^j + \Phi^i (x)$ in such a way that $\pa \Phi^i / \pa x^j = 0$ for $j > i$. It is then possible to solve the equations recursively, starting from the linear one for $x^1 (t)$ and having at each step a linear equation with a forcing term which is a nonlinear but explicitely known function of $t$. 

The procedure proposed here is equivalent to the one considered by  Dulac (and attributed by him to Horn and Lyapunov) from the analytic point of view\footnote{Note it can be more convenient computationally, as it only requires to solve linear systems; in particular it will be conveniently implemented on computers via algebraic manipulation languages.}, but it has the advantage of showing how the nonlinear (normal form) system is obtained by restrictrion (on the submanifold $\M$) of a linear system via nonlinear constraints, clarifying the geometry involved in the integrability and integration of resonant normal forms with Poincar\'e spectrum, and the connection with topics in modern integrable systems theory \cite{KKS,Mar}. It also shows that these ideas, developed in the hamiltonian context, can be fruitfully  applied to more general dynamical systems.

\section{Examples}

Systems in normal form are specially interesting if a small neighbourhood of the origin is dynamically invariant, i.e. if the critical point is stable (in this case the evolution will remain in the domain of analyticity of the normalizing transformation); thus we are mainly interested in cases where the real part of the eigenvalues is negative (or zero). However, for ease of notation we will consider examples with positive eigenvalues; the stable situation is recovered by a time reversal. Also for ease of notation, we will write all vector indices as lower ones; the $c_i$ will be arbitrary real constants.

\subsection*{Example 1.} 

Let us consider $n=2$, with coordinates $(x,y)$ and 
$$ A \ = \ \pmatrix{1&0\cr0&k\cr}  $$
with $k$ a positive integer; notice here $\s = \{1,k\}$ is in a Poincar\'e domain.
There is only one resonance $\mu = (k,0)$ (with $\a =2$), and the only resonant monomial is $\phi (x , y )  = x^k$. The Poincar\'e-Dulac normal form is
$$ \begin{array}{ll}
{\dot x} = & x \\ 
{\dot y} = & k y + c_1 x^k \end{array} $$ 
with $c_1$ an arbitrary coefficient.

Thus, following our procedure, we set ${\dot w} = k (x^{k-1}) {\dot x} = k w$; the system obtained in $V$ is
$$ \begin{array}{ll} 
{\dot x} = & x \\ 
{\dot y} = & k y + c_1 w \\ 
{\dot w} = & k w \end{array}$$
and the constraint $\E$ is given by $w - x^k = 0$.

The solution to the system is given by 
$$ x(t) = x_0 e^t \ , \ y(t) = y_0 e^{kt} + (c_1 k w_0) t e^{kt} \ , \ w(t) = w_0 e^{kt} \ ; $$
obviously the submanifold $\M$ identified by $w = x^k$ is invariant under this flow, and the projection of solutions on $\M$ to $\R^2 = (x,y)$ is simply
$$ x(t) \ = \ x_0 e^t \ \ , \ \ y (t) \ = \ [ y_0 + (c_1 k x_0^k) t ] \, e^{kt} $$

\subsection*{Example 2.}

Let us consider $n=3$, with coordinates $(x,y,z)$ and
$$ A \ = \ \pmatrix{1&0&0\cr0&2&0\cr0&0&5\cr} \ ; $$
notice here $\s = \{1,2,5\}$ is in a Poincar\'e domain.
There are four resonances: 
$$ \cases{
\mu = (2,0,0) & (with $\a =2$ and $|\mu | = 2$) \cr
\mu = (1,2,0) & (with $\a =3$ and $|\mu | = 3$) \cr
\mu = (3,1,0) & (with $\a =3$ and $|\mu | = 4$) \cr
\mu = (5,0,0) & (with $\a =3$ and $|\mu | = 5$) \cr } $$
and correspondingly we have 
$$ \phi_1  = x^2 \ , \ \phi_2 = x y^2 \ , \ \phi_3 = x^3 y \ , \ \phi_4 = x^5 \ . $$
The normal form is written as 
$$ \begin{array}{ll}
{\dot x} =& x \\ 
{\dot y} =& 2 y + c_1 x^2 \\ 
{\dot z} =& 5 z + c_2 x y^2 + c_3 x^3 y + c_4 x^5 \end{array}$$
with $c_i$ arbitrary real coefficients.

By our procedure, we obtain a system in $V = \R^7$, given by 
$$ \begin{array}{ll} 
{\dot x} = & x \\ 
{\dot y} = & 2 y + c_1 w_1 \\ 
{\dot z} = & 5 z + c_2 w_2 + c_3 w_3 + c_4 w_4 \\
{\dot w}_1 =& 2 w_1 \\ 
{\dot w}_2 =& 5 w_2 + 2 c_1 w_3 \\ 
{\dot w}_3 =& 5 w_3 + c_1 w_5 \\ 
{\dot w}_4 =& 5 w_4 \end{array} $$
(notice this has the block structure discussed in section 2). 
The general solution to this (writing $w_i (0) = \ga_i$ for ease of notation) is 
$$ \begin{array}{l}
x(t) = x_0 e^t \ \ \ \ , \ \ \ \ y(t) = (y_0 + \ga_1 t ) e^{2t} \\ 
z(t) = \[ z_0 + (\ga_2 c_2 + \ga_3 c_3 + \ga_4 c_4 ) t + (1/2) (2 \ga_3 c_1 c_2 + \ga_4 c_3 ) t^2 \] e^{5t} \\ 
w^1 (t) = \ga_1 e^{2t} \ \ \ \ , \ \ \ \ w^2 (t) = (\ga_2 + 2 c_1 \ga_3 t ) e^{5t} \\
w^3 (t) = ( \ga_3 + \ga_4 t ) e^{5t} \ \ \ \ , \ \ \ \ w^4 (t) = \ga_4 e^{5t} \end{array} $$
which once restricted to the manifold $\M$ (here identified by 
$\E_1 := w_1 - x^2$, $\E_2 := w_2 - x y^2$, $\E_3 := w_3 - x^3 y$, $\E_4 := w_4 - x^5$) and projected to $\R^n$, gives
$$ \begin{array}{l}
x(t) \ = \ x_0 \ e^t \\
y(t) \ = \ (y_0 \, + \, x_0^2 \, t ) \ e^{2t} \\ 
z(t) \ = \ \[ z_0 \, + \, (c_2 x_0 y_0^2 + c_3 x_0^3 y_0 + c_4 x_0^5 ) \, t \, + \,  (2 c_1 c_2 x_0^3 y_0 + c_3 x_0^5 ) \, (t^2 /2)  \] \  e^{5t} \end{array} $$

\subsection*{Example 3.}

Let us consider $n=3$, with coordinates $(x,y,z)$ and
$$ A \ = \ \pmatrix{1&\eps&0\cr0&1&0\cr0&0&2\cr} \ ; $$
notice here $\s = \{1,1,2\}$ is in a Poincar\'e domain, and $A_n \not=0$ for $\eps \not= 0$.

There are three resonances, all of them with $\a = 3$ and $|\mu | = 2$, i.e.
$$  \mu = (2,0,0) \ , \ \mu = (1,1,0) \ , \ \mu = (0,2,0) $$
and correspondingly
$$ \phi_1 = x^2 \ , \ \phi_2 = x y \ , \ \phi_3 = y^2 \ . $$
The seminormal form with respect to the semisimple part $A_s$ of $A$ is thus
$$ \begin{array}{ll}
{\dot x} =& x + \eps x \\ 
{\dot y} =& y \\ 
{\dot z} =& 2 z + c_1 x^2 + c_2 x y + c_3 \end{array} $$
The normal form with respect to the full $A$ is the same for $\eps = 0$ (which means $A = A_s$), and for $\eps \not=0$ is obtained setting $c_2 = c_3 = 0$, as readily seen by considering $X_F = (c_1 x^2 + c_2 xy + c_3 y^2 ) \pa_z$ and imposing $[X_\ell , X_F ] = 0$.

The associated linear system in $V=\R^6$ is $ {\dot \xi} = B \xi $, with 
$$ B \ = \ \ \pmatrix{
 1 & 0 & 0 & 0   & 0   & 0   \cr
 \eps & 1 & 0 & 0 & 0 & 0 \cr
 0 & 0 & 2 & c_1 & c_2 & c_3 \cr
 0 & 0 & 0 & 2 & \eps & 0 \cr
 0 & 0 & 0 & 0 & 2 & \eps \cr
 0 & 0 & 0 & 0 & 0 & 2 \cr}  $$
for the seminormal form, and the same -- with $c_2 = c_3 = 0$ if $\eps$ is nonzero -- when we normalize with respect to the full $A$.

Writing again $w_i (0) = \ga_i$, the general solution to this linear system is 
$$ \begin{array}{l}
x(t) = x_0 e^t \\ 
y(t) = (y_0 + \eps x_0 t) e^t \\
z(t) = \[ z_0 + (c_1 \ga_1 + c_2 \ga_2 + c_3 \ga_3 ) t + (c_1 \ga_2 + \eps c_2 \ga_3 ) (t^2 /2) + \eps c_1 \ga_3 (t^3 /6) \] e^{2t} \\ 
w_1 (t) = (\ga_1 + \ga_2 t + \eps \ga_3 t^2/2 ) e^{2t} \\
w_2 (t) = (\ga_2 + \eps \ga_3 t ) e^{2t} \\  
w_3 (t) = \ga_3 e^{2t} \end{array} $$
Restricting to $\M$ (which in this case is identified by $\E_1 := w_1 - x^2$, $\E_2 := w_2 - x y$, $\E_3 := w_3 - y^2$) and projecting to $\R^3$ gives
$$ \begin{array}{ll}
x(t) \ =& \ x_0 \ e^t \\ 
y(t) \ =& \ (y_0 \, + \, \eps x_0 \, t ) \ e^t \\
z(t) \ =&  \ [ z_0 \, + \, (c_1 x_0^2 + c_2 x_0 y_0 + c_3 y_0^2 ) \, t \ + \\ 
& \ \ + \ (c_1 x_0 y_0 +  \eps c_2 y_0^2 ) \, (t^2 /2) \, + \, \eps c_1 y_0^2 \, (t^3 /6) ] \ e^{2t} \end{array} $$

If $\eta \not= 0$ and the system is in normal form with respect to the full matrix $A$ (so that $c_2 = c_3 = 0$) then $z(t)$ simplifies to 
$$ z(t) \ = \ \[ z_0 + c_1 x_0^2 t + c_1 x_0 y_0 (t^2 /2) + \eps c_1 y_0^2 (t^3 /6) \] e^{2t} \ . $$

Notice that (for $\eta \not= 0$) the time evolution of $\phi_1$ depends on $\phi_2$, and through this on $\phi_3$ as well; thus, it would not be possible to consider a parent system involving only terms in normal form with respect to $A^+$. For $\eta = 0$ the equations for the $w_i$ decouple, but then all the $\phi_i$ are allowed in the normal form.

\subsection*{Example 4.}

Let us consider $n=2$ with coordinates $(x,y)$ and
$$ A \ = \ \pmatrix{0&-1\cr 1&0\cr} $$
Here $\s = \{ - i , i \}$ does not satisfy the Poincar\'e condition. We have a master resonance $\la_1 + \la_2 = 0$, and hence an infinite number of resonances, given by 
$$ \cases{\mu = (k+1,k) & i.e. $(k + 1) \la_1 + k \la_2 = \la_1$ , $\a = 1$ \cr
\mu = (k , k+1) & i.e. $k \la_1 + (k + 1) \la_2 = \la_2$ , $\a = 2$ \cr} $$
and the resonant monomials are given by 
$$ \phi_{2m-1} = (x^2 + y^2)^m x = \rho^m x \ , \ \phi_{2m} = (x^2 + y^2)^m y = \rho^m y\ , $$
where $\rho = (x^2 + y^2 )$. For ease of notation we introduce coordinates $q_m = w_{2m-1} = \phi_{2m-1}$, $p_m = w_{2m} = \phi_{2m}$.

The normal form is then 
$$ \begin{array}{ll}
{\dot x} =& - y + \sum_{k=1}^\infty  (x^2 + y^2)^k (a_k x - b_k y) \\ 
{\dot y} =& + x + \sum_{k=1}^\infty  (x^2 + y^2)^k (b_k x + a_k y) \end{array} $$ 
and this is rewritten following our procedure as an infinite linear system: 
$$ \begin{array}{ll}
{\dot x} =& - y + \sum_{k=1}^\infty   (a_k q_k - b_k p_k ) \\ 
{\dot y} =& + x + \sum_{k=1}^\infty   (b_k q_k + a_k p_k ) \\ 
{\dot q}_k =& - p_k + \sum_{s=1}^\infty \ \[ (2k +1) a_s q_{k+s} - b_s p_{k+s} \] \\ 
{\dot p}_k =& + q_k + \sum_{s=1}^\infty \ \[  b_s q_{k+s} + (2k +1) a_s p_{k+s} \] \end{array} $$

Notice that truncating this at order $N = 2k +1$, i.e. projecting it down to the linear subspace of $(x,y)$ and the $(p_m,q_m)$ with $m \le k$, we get a finite dimensional linear system (all sums go then from $1$ to $k$).
However, the dynamics of this projected system can and in general does fail to reproduce even qualitatively the full dynamics; actually, it will also not preserve the manifold $\M$. 
To see this, it suffices to consider the truncation at $k=1$, $N=3$, which reads
$$ \begin{array}{ll}
{\dot x} =& - y + (a_1 q_1 - b_1 p_1 ) \\
{\dot y} =& + x + (b_1 q_1 + a_1 p_1 ) \\
{\dot q}_1 =& - p_1 \\
{\dot p}_1 =& + q_1 \ . \end{array} $$

\subsection*{Example 5.}

Let us briefly consider the situation mentioned in subsection 2.3; let  $n=4$ with coordinates $x_i$ ($i=1,...,4$) and 
$$ A \ = \ {\rm diag} \ (1,2,3,10) \ ; $$
here again $\s$ is a Poincar\'e spectrum. We want to consider normal forms truncated at order $|x|^3$; the resonances of order $|\mu| \le 3$ are given by 
$$ \cases{ 
\mu = (2,0,0,0) & (with $\a = 2$ and $|\mu|=2$) \cr
\mu = (1,1,0,0) & (with $\a = 3$ and $|\mu|=2$) \cr
\mu = (3,0,0,0) & (with $\a = 3$ and $|\mu|=3$) \cr} $$
There are then several resonances of higher order, with $4 \le |\mu| \le 10$, all of them with $\a = 4$. 

We will introduce 
$$ \phi_1 = x_1^2 \ , \ \phi_2 = x_1 x_2 \ , \ \phi_3 = x_1^3 \ . $$
The normal form corresponding to this linear part $A$ is
$$ \begin{array}{rl}
{\dot x}_1 \ = \ & x_1 \\
{\dot x}_2 \ = \ & 2 \, x_2 \ + \ c_1 \, x_1^2 \\
{\dot x}_3 \ = \ & 3 \, x_3 \ + \ c_2 \, x_1 x_2 \ + \ c_3 \, x_1^3 \\
{\dot x}_1 \ = \ & 10 \, x_4 \ + \ O (|x|^4) \end{array} $$
where the $c_i$ are real constants, and the truncated normal form of order three is obtained by dropping the term $O(|x|^4)$ in the above. 

By applying our procedure to this {\it truncated} normal form we get
$$ \begin{array}{rl}
{\dot x}_1 \ = \ & x_1 \\
{\dot x}_2 \ = \ & 2 \, x_2 \ + \ c_1 \, \phi_1 \\
{\dot x}_3 \ = \ & 3 \, x_3 \ + \ c_2 \, \phi_2 \ + \ c_3 \, \phi_3 \\
{\dot x}_1 \ = \ & 10 \, x_4 \\
{\dot \phi}_1 \ = \ & 2 \, \phi_1 \\
{\dot \phi}_2 \ = \ & 3 \, \phi_2 \ + \ c_1 \, \phi_3 \\
{\dot \phi}_1 \ = \ & 3 \, \phi_3 \end{array} $$
We stress that this is not a truncation of the obtained equations: once we deal with the truncated normal form, we obtain exactly this closed form (indeed, no $\phi$ depends on $x_4$), as shown in subsection 2.3.

\vfill\eject

\end{document}